\documentclass[aps,pra,twocolumn,showpacs,preprintnumbers,amsmath,amssymb,nofootinbib,superscriptaddress]{revtex4-1}
\usepackage{mathtools}
\usepackage[pdftex,
            pdfauthor={Tom Burkart, Manon C. Wigbers, Laeschkir Wuerthner, and Erwin Frey},
            pdftitle={Control of protein-based pattern formation via guiding cues},
            pdfsubject={biorXiv pre-print},
            urlcolor=blue,
            linkcolor=blue,
            citecolor=blue,
            colorlinks]{hyperref}
\usepackage[utf8]{inputenc}
\usepackage{graphicx}

\begin{document}

\title{Control of protein-based pattern formation via guiding cues}

\author{Tom Burkart}
\thanks{These authors contributed equally.}
\affiliation{Arnold Sommerfeld Center for Theoretical Physics (ASC) and Center for NanoScience (CeNS), Department of Physics, Ludwig-Maximilians-Universit\"at M\"unchen, Theresienstra{\ss}e 37, D-80333 Munich, Germany}

\author{Manon C. Wigbers}
\thanks{These authors contributed equally.}
\affiliation{Arnold Sommerfeld Center for Theoretical Physics (ASC) and Center for NanoScience (CeNS), Department of Physics, Ludwig-Maximilians-Universit\"at M\"unchen, Theresienstra{\ss}e 37, D-80333 Munich, Germany}

\author{Laeschkir W\"urthner}
\thanks{These authors contributed equally.}
\affiliation{Arnold Sommerfeld Center for Theoretical Physics (ASC) and Center for NanoScience (CeNS), Department of Physics, Ludwig-Maximilians-Universit\"at M\"unchen, Theresienstra{\ss}e 37, D-80333 Munich, Germany}

\author{Erwin Frey}
\email{frey@lmu.de}
\affiliation{Arnold Sommerfeld Center for Theoretical Physics (ASC) and Center for NanoScience (CeNS), Department of Physics, Ludwig-Maximilians-Universit\"at M\"unchen, Theresienstra{\ss}e 37, D-80333 Munich, Germany}
\affiliation{Max Planck School Matter to Life, Hofgartenstra{\ss}e 8, D-80539 Munich, Germany}

\date{February 11, 2022}

%
%
%
%

\begin{abstract}
Proteins control many vital functions in living cells, such as cell growth and cell division.
Reliable coordination of these functions requires the spatial and temporal organizaton of proteins inside cells,
which encodes information about the cell's geometry and the cell-cycle stage.
Such protein patterns arise from protein transport and reaction kinetics, and they can be controlled by various guiding cues within the cell.
Here, we review how protein patterns are guided by cell size and shape, by other protein patterns that act as templates, and by the mechanical properties of the cell.
The basic mechanisms of guided pattern formation are elucidated with reference to recent observations in various biological model organisms.
We posit that understanding the controlled formation of protein patterns in cells will be an essential part of understanding information processing in living systems.
\end{abstract}

\maketitle%
%
%
%
%
\section{Introduction}
To ensure their survival, cells must tightly regulate a wide range of cellular functions, such as cell migration, cell growth, DNA synthesis, and cell division. 
For example, in order to produce two viable daughter cells, a cell must precisely coordinate cell growth with the duplication and segregation of DNA, and with subsequent cell division.
These cellular functions, in turn, are controlled and coordinated by proteins.
Robust timing and reliable control of these functions requires cells to process \emph{spatiotemporal information}, such as information about cell size and shape, cell cycle state, the cell's surroundings, and the current state of other cellular processes.
Such spatiotemporal information is encoded in \emph{protein patterns} -- i.e.,~an inhomogeneous spatial distribution of proteins -- that regulate these cellular functions, whereby each type of protein may perform distinct tasks.

How then are proteins spatially and temporally organized in a cell?
The idea that the collective organization of interacting chemicals (chemical reactions) in an initially homogeneous medium can give rise to spatial patterns dates back to Turing's seminal work on spontaneous pattern formation in reaction-diffusion systems~\cite{Turing1952}.
While this work has greatly advanced the understanding of pattern formation in biological systems, many aspects of protein patterns such as their positioning, timing, reliability, and controllability -- which are essential for the viability of living organisms -- remain poorly understood.
Since protein patterns in cells serve a timed and targeted functional purpose, they must form in response to certain signals and control mechanisms rather than spontaneously emerging from an initially homogeneous distribution.
Indeed, an increasing number of theoretical and experimental studies find that protein distributions can respond and adapt to cell shape, size, and mechanics, as well as to signals encoded in previously established protein patterns~\cite{Thalmeier.etal2016, Gessele.etal2020, Gross.etal2019, Haupt.Minc2018, Moseley.Nurse2010, Goychuk.Frey2019, Hubatsch.etal2019}.

This response is, in fact, bidirectional.
Cells are not static objects but rather an active material whose size, shape, and mechanical properties can be altered dynamically through protein interactions in response to the cell's environment and the current state of the cell cycle~\cite{Kasza.etal2007, Lecuit.Lenne2007, Mierke2020, Cadart.etal2019}.
These dynamic interactions between protein patterns and cell architecture are the subject of a rapidly developing field of study at the interface between cell biology and theoretical physics that benefits from constantly improving experimental techniques, as well as insights from physics that allow one to model and understand the guided organization of proteins into patterns.

In this review, we summarize recent advances in our understanding of how protein patterns are controlled by geometric, mechanical, and biochemical cues.
The basics of pattern formation will only be summarized briefly, as recent reviews have provided a comprehensive introduction to this subject.
The interested reader is referred to an elementary course on the mathematical tools that are required to study the physics of protein interactions and pattern formation, in particular ordinary differential equations (ODEs) and nonlinear dynamics~\cite{Strogatz1994}.
For an introduction to the theory of pattern-forming systems, we direct the reader to pertinent textbooks~\cite{Cross.Greenside2009, Desai.Kapral2009}, and to lecture notes for a review on quantitative modeling of pattern formation in mass-conserving systems~\cite{Frey.Brauns2020}.
Other recent reviews have focused on the theory of two specific aspects of pattern formation, namely the role of bistability for polarity~\cite{Champneys.etal2021}
and the curvature-generating properties of proteins~\cite{Alimohamadi.Rangamani2018}.
The relevance of protein patterns for cells has also been reviewed from a more biological perspective recently~\cite{Shapiro.etal2009}, in particular with respect to midcell localization~\cite{Lutkenhaus2012}, and current advances in understanding pattern formation at a molecular level~\cite{Kretschmer.etal2018} have been reviewed recently.
We also want to highlight three recent reviews that emphasize the importance and role of modeling for understanding cell polarity~\cite{Edelstein-Keshet.etal2013, Goryachev.Leda2020} and biological phenomena in general~\cite{Moebius.Laan2015}.

Here, we discuss several theoretical models that have been developed with a view to reproducing and accounting for pattern guidance, together with examples of well-studied biological model organisms in which pattern guidance has been observed to play a critical role in cell viability.
In particular, we discuss how biophysical theory has been instrumental in clarifying the underlying physical concepts of pattern guidance in living cells.
We start by giving an overview of the predominant types of protein transport and chemical reactions that are predominately involved in the formation of patterns in cells.
We then discuss how these factors can be affected by cell shape and size, pre-existing protein patterns, and cell mechanics, and how these cues guide and control protein pattern formation.
We conclude with an outlook on the future research directions in this field.

\section{Basic principles of pattern formation}%
\begin{figure*}
    \centering
    \includegraphics[width = 0.8\textwidth]{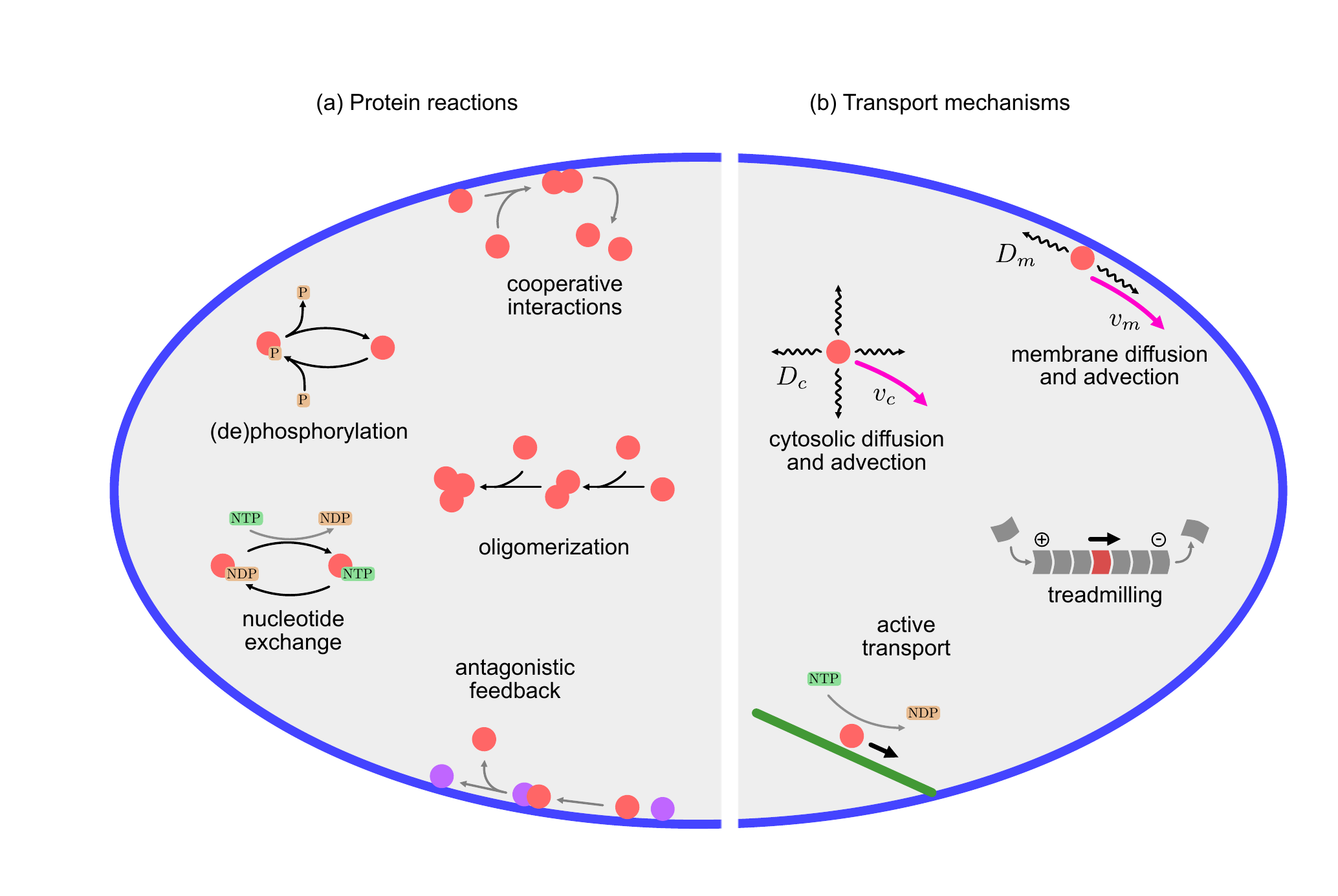}
    \caption{\textbf{Reaction and transport processes involved in pattern formation:}
    (a) Protein reactions 
    include binding to and detachment from the cell membrane or other intracellular structures, as well as conformational state changes due to (de-)phosphorylation or nucleotide exchange.
    Cooperative and antagonistic (nonlinear) reactions between multiple proteins can lead to assisted attachment (recruitment) or to detachment from the membrane.
    Multiple monomers can form oligomers with altered transport and reaction properties.
    (b) Proteins can be transported by diffusion ($D_c,\, D_m$, black arrows) and advection ($v_c,\, v_m$, pink) independently on surfaces -- in particular cell cortex and membrane -- and in the cytosol. In addition, directed protein transport can be established by subunit addition and disassembly of polymers, resulting in treadmilling of monomers, and by active transport along filamentous structures, mediated by energy-consuming motor proteins.
    }
    \label{fig:transport_reactions}
\end{figure*}%
Protein patterns arise from the interplay of biochemical reaction kinetics with different types of transport mechanisms. 
While the amounts of locally available proteins are regulated by chemical reactions, their spatial distribution is altered by transport processes including diffusion, active transport and fluid flow (see Fig.~\ref{fig:transport_reactions}).
Some of the most important reaction and transport processes involved are presented in the following.

\subsection{Protein reaction networks}%
Protein reaction networks differ in their degree of complexity, e.g., with respect to the number of different proteins and their conformations, as well as the number and type of reactions between them.
Some of the most common types relevant to protein pattern formation are briefly discussed in the following.

\textit{Conformational state changes} --
The intracellular organization of proteins is largely controlled by protein reaction networks that contain nucleoside triphosphate\footnote{Nucleoside tri-/diphosphate (NTP/NDP) -- Nucleotide molecules with three (two) phosphate groups typically based on guanine (GTP), adenine (ATP) or cytosine (CTP), forming the main carriers of chemical energy in cells.} (NTP)-dependent regulatory modules.
In prokaryotic cells, P-loop (phosphate binding loop) ATPases\footnote{NTPase -- Enzymes that bind to NTP and hydrolize it to NDP, thereby releasing energy.} such as ParA and MinD take on this role, and give rise to self-organized dynamic patterns at cellular interfaces -- ParA on the nucleoid and MinD on the cell membrane~\cite{Lutkenhaus2012, Bange.Sinning2013, Vecchiarelli.etal2012}.
Similarly, small \mbox{GTPases} like Cdc42 and RhoA play an important role in establishing cell polarity in eukaryotic cells~\cite{Iden.Collard2008, Etienne-Manneville2004, Perez.Rincon2010}.
Basically, all these proteins serve as molecular switches that can cycle between an active and inactive state based on nucleotide binding and delayed hydrolysis, typically regulated by auxiliary proteins~\cite{Bokoch.etal1994,Ubersax.FerrellJr2007,Irazoqui.etal2003} (Fig.~\ref{fig:transport_reactions}a).
Similarly, proteins that are not NTPases can act as molecular switches if cycling between active and inactive states (phosphorylation\footnote{Phosphorylation -- Proteins can be (de-)phosphorylated by the addition of a phosphate group, as a means of storing (releasing) chemical energy.} and dephosphorylation) is catalyzed by separate kinases and phosphatases, respectively~\cite{Kuo.etal2014, Hoege.Hyman2013}.
These cycles have two key features.
First, they are non-equilibrium processes driven by the supply of chemical energy, e.g.~through ATP hydrolysis~\cite{Alberts.etal2002}.
As such, they are the core element of most protein reaction networks, enabling them to drive self-organization processes.
Secondly, the switch between active and inactive states is associated with changes in their affinity for targets such as the cell membrane and the nucleoid~\cite{Alberts.etal2002, Osorio-Valeriano.etal2019}, as well as their specific binding affinity for other proteins or lipids.
For example, MinD can only bind to the cell membrane in its ATP-bound, dimeric form and is released into the cytosol as an ADP-bound monomer upon ATP hydrolysis~\cite{Lackner.etal2003}.

\textit{Binding and unbinding reactions} -- 
Many proteins can bind to different substrates~in a cell, such as membranes.
Typical residence times of proteins on membranes range from seconds to minutes~\cite{Goehring.etal2011a, Robin.etal2014, Gross.etal2019}.
In several biological model systems, the nonlinear binding kinetics of proteins to membranes plays a key role in the formation of spatiotemporal protein patterns.

One way to confer nonlinear binding kinetics is through limitation of binding sites on the membrane, which leads to saturated binding kinetics~\cite{Goryachev.Leda2017}.
Another example is cooperative reactions that amplify or attenuate the attachment and detachment of other proteins to the membrane~\cite{Ramm.etal2019, Halatek.etal2018, Goryachev.Leda2017a, Heermann.etal2021} (Fig.~\ref{fig:transport_reactions}a).
These feedback mechanisms were shown to be an integral part of the patterning mechanisms in the most important model organisms:
In the MinDE system of \textit{E.~coli}, pole-to-pole oscillations of the Min proteins rely on recruitment of cytosolic MinD and MinE by membrane-bound, active MinD (positive feedback) and their release into the cytosol through MinE-induced hydrolysis and concomitant inactivation of MinD (negative feedback)~\cite{Hu.Lutkenhaus2001, Lackner.etal2003, Miyagi.etal2018, Halatek.Frey2012}.
In budding yeast (\textit{S.~cervisiae}), the establishment of cell polarity via asymmetric distribution of Cdc42 involves multiple positive and negative feedback loops, which provide a high degree of robustness~\cite{Irazoqui.etal2003, Howell.etal2012, Goryachev.Leda2017, Brauns.etal2020a}.
Finally, the PAR polarity system in the early \textit{C.~elegans} embryo exploits various antagonistic reactions that play a key role in specifying the correct orientation of the polarity axis~\cite{Hoege.Hyman2013, Motegi.etal2011, Munro.etal2004, Hao.etal2006}.

\textit{Complex formation} --
Proteins can also form oligomers\footnote{Oligomer -- Complex made up of a few proteins of the same or or a different type (homo- and hetero-oligomers, respectively).}, in particular dimers~(Fig.~\ref{fig:transport_reactions}a). 
This can have an impact on their ability to bind to cellular surfaces, as described above for active MinD dimers.
The formation of higher-order protein aggregates leads to a change in P{\'e}clet number (see below), which in turn alters how they are affected by fluid flow as opposed to diffusion.
Such an effect has been suggested to play a role in the transport of PAR-3 proteins in the \textit{C.~elegans} embryo.
Here, diffusive transport may dominate for PAR-3 monomers (${\textrm{Pe}<1}$), whereas transport becomes dominated by flow (${\textrm{Pe}>1}$) upon cell-cycle-dependent aggregation of PAR-3 into complexes together with two other proteins -- PAR-6 and aPKC~\cite{Gubieda.etal2020}.
Yet another process is the formation of higher-order oligomers, such as those observed for membrane-bound MinD~\cite{Miyagi.etal2018, Heermann.etal2021}.
Similar to the nonlinear attachment kinetics discussed above, cooperative reactions have also been suggested to participate in protein complex formation, potentially allowing for feedback loops~\cite{Hoege.Hyman2013}.

\textit{Theory --}
Mathematically, the dynamics of well-mixed protein reaction networks are described by sets of coupled nonlinear differential equations for the concentrations $u_i (t)$ of each of the different protein types and conformations ${i \in \{1, \ldots, S\}}$,
\begin{equation}
    \partial_t u_i (t) 
    =
    f_i (\{ u_i \}) \, .
\end{equation}
In such \emph{chemical rate equations}, the nonlinear reaction terms $f_i$ (together with the reaction rates) must be inferred from the underlying reaction network using the law of mass action.
An elaborate mathematical theory, called \textit{dynamic system theory}, allows one to analyze systems of coupled nonlinear ordinary differential equations (ODEs). 
The basic idea of this theory, which goes back to the pioneering work of Poincar{\'e}~\cite{Poincare1993}, is to characterize the system dynamics in terms of certain geometric structures in the phase space spanned by the set of dynamical variables $u_i (t)$~\cite{Strogatz1994,Cross.Greenside2009}.

Of particular interest are the asymptotic dynamics of the system over large time scales, which are characterized by the \emph{attractors} in phase space within the framework of dynamic system theory.
These include fixed points corresponding to reactive equilibria (see Supplementary Information), limit cycles corresponding to nonlinear oscillators, and more intricate geometric objects~\cite{Strogatz1994,Cross.Greenside2009}.
Importantly, the local properties of the fixed points (reactive equilibria), in particular their stability, can be determined using ODEs linearized around these fixed points~\cite{Strogatz1994,Cross.Greenside2009}.

\subsection{Protein transport}%
\label{sec:transport}%
Transport mechanisms play a crucial role in the control of spatial variations in protein concentration. 
In the following, we provide an overview of the most important modes of intracellular protein transport involved in pattern formation (Fig.~\ref{fig:transport_reactions}b).

\textit{Diffusion} --
Perhaps the most basic means of protein transport is diffusion. 
It is a consequence of Brownian motion and is directed from regions of high to regions of low protein concentration $u({\bf x},t)$ with a diffusive current ${- D \nabla u({\bf x},t)}$ (Fick's law). 
For spherical particles of radius $r$, the diffusion constant is given by the Stokes-Einstein relation ${D= k_B T/(6 \pi \eta r)}$, where $\eta$ is the viscosity of the surrounding cytosol~\cite{Frey.Kroy2005}; a qualitatively similar relation holds for transmembrane proteins~\cite{Saffman.Delbrueck1975, Petrov.Schwille2008}.
This implies that the diffusive transport of proteins depends on their size and on the local properties of the surrounding medium.

Importantly, both the membrane and the cytoplasm\footnote{Cytoplasm -- Heterogeneous material making up most of the volume of a cell (excluding the nucleus), mainly consisting of the cytosol and macromolecular organelles.} are highly heterogeneous environments crowded with macromolecular structures that interact with proteins, for example by temporarily binding or by taking up space~\cite{Agrawal.etal2022}.
For the purpose of studying pattern formation, however, one often disregards inhomogeneities and instead assumes an effective diffusion constant that takes into account such interactions that are not explicitly modeled.
Hence, the diffusion constant is a mesoscopic quantity representing the mobility of proteins in a homogeneous, dilute fluid environment.
In essence, the complex cytoplasmic environment is reduced to an effective cytosol for many applications in protein pattern formation, and similarly, the heterogeneous membrane is considered as an effective (dilute) fluid~\cite{Hoefling.Franosch2013}.
This simplification is justified since the length scale of patterns is typically larger than the length scale of heterogeneities in the cytoplasm or on the membrane, to which we will refer as substrates in the following.
As a rough estimate, the diffusion coefficients of membrane-bound proteins are generically at least two orders of magnitude lower than those of their cytosolic counterparts: While characteristic values for membrane diffusion are $D_m \sim 0.01 \, \mu m^2/s$, one observes $D_c \sim 10 \, \mu m^2/s$ in the cytosol~\cite{Meacci.etal2006}.
Although the models discussed in this review suggest that the heterogeneous character of the cellular substrates are of minor importance for protein pattern formation, it would be interesting to explicitly probe the robustness of these models against more realistic substrates.
For example, this could be incorporated into models via time- and space-dependent diffusion constants.

\textit{Active transport} --
Proteins can also be transported via active processes driven by the chemical energy of ATP, GTP or CTP at the molecular level. 
Of particular biological relevance are translational molecular motors\footnote{Molecular motors -- Enzymes that use energy released by NTP hydrolysis to perform mechanical work and that are generally associated with cytoskeletal filaments.}~\cite{Vale2003, Schliwa.Woehlke2003, Kolomeisky2015}.
An important subclass of these motors is comprised of kinesins and dyneins that bind to, and ‘walk’ on microtubules\footnote{Microtubules and actin filaments -- Protein filaments comprised of tubulin and actin proteins, respectively, which form an integral part of the cytoskeleton.}.
In this way, cargo -- such as other proteins -- can be transported along the microtubules~\cite{Vale2003, Schliwa.Woehlke2003}. 
Depending on the type of motor and, in some cases, other factors such as external forces~\cite{Pandey.etal2021}, this form of active transport is directed to either the plus or minus end of the microtubules~\cite{Woehlke.Schliwa2000}.
Certain classes of myosin motors perform similar tasks by transporting cargo along actin filaments.
Such active cargo transport is known to be involved in the polarization process of budding yeast.
Here, the actin filaments are anchored to the polarity site, so that the myosin motors can deliver protein-coated vesicles towards the polarity site~\cite{Jin.etal2011, Evangelista.etal2002}.

Another class of active transport processes is mediated by the directed polymerization of cytoskeletal filaments such as F-actin~\cite{Mogilner.Oster2003} and microtubules~\cite{Desai.Mitchison1997}, which is driven by ATP and GTP hydrolysis, respectively. 
For instance, tubulin-like FtsZ filaments are particularly important active structures in bacterial cell division.
These filaments exhibit treadmilling dynamics (see the segmented structure in Fig.~\ref{fig:transport_reactions}b), as FtsZ monomers can only bind to the plus end and detach from the minus end~\cite{Stricker.etal2002, Loose.Mitchison2014}.
By consuming GTP, this treadmilling allows FtsZ filaments to translocate directionally along the cell membrane, coordinating the activity of downstream cell division processes~\cite{Bisson-Filho.etal2017}.
Similarly, treadmilling of actin filaments was shown to play a key role in cell migration, in particular for the extrusion of lamellipodia~\cite{Krause.Gautreau2014}.

Both \textit{in vivo} and \textit{in vitro} experiments have shown how important these active transport processes are for the polarization of cells~\cite{Snaith.etal2005, Minc.etal2009, Gennerich.Vale2009, Langford2002, Mata.Nurse1997}.
For example, during cell growth in fission yeast microtubules are aligned along the long axis of the cell, and direct the active transport of the tip factors Tea1 and Tea4 towards the cell poles in a two-fold manner~\cite{Chiou.etal2016, Huisman.Brunner2011, Mata.Nurse1997, Tatebe.etal2005}:
The kinesin-like motor Tea2 mediates the transport of Tea1/Tea4 complexes along microtubules that emanate from the nucleus~\cite{Browning.etal2000, Vendel.etal2019}.
In addition, these complexes bind to microtubule tips assisted by Mal3, a tip-binding protein.
Therefore, due to the directed microtubule polymerization along the long cell axis, the tip factors are transported to the cell poles~\cite{Vendel.etal2019}.
At the poles, they then serve as a spatial cue for cell growth, and therefore facilitate the elongation of the cell along its long axis~\cite{Tay.etal2018}.

\textit{Advective transport} -- In the fluid environment of a cell, proteins can also be transported by cytoplasmic~\cite{Goldstein.Meent2015, Vecchiarelli.etal2014a}, cortical~\cite{Munro.etal2004}, and membrane flows~\cite{Gerganova.etal2021, Vecchiarelli.etal2016}, whose effect on protein transport through friction strongly depends -- like diffusion -- on the viscosity of the respective environment.
An important force-generating active structure is the actin cortex\footnote{Actin cortex -- Thin and dynamic network that acts as a scaffold that determines the cell's shape and which is comprised of actin filaments, motor proteins, and other associated proteins.}.
In addition to actin filaments, it includes cross-linker proteins and myosin motors that cause cortical contractions which, in turn, can induce flows~\cite{Salbreux.etal2012, Chugh.Paluch2018}.
The cortical contractions that occur in the \emph{C.~elegans} zygote are a prominent example~\cite{Goehring.etal2011, Gross.etal2019, Illukkumbura.etal2020}.
Here, local depletion of the concentration of the motor protein myosin at the cell cortex leads to a gradient of contractile stress, such that the cell cortex flows from the anterior to the posterior pole~\cite{Klinkert.etal2019}.

Cortical contractions can also lead to flows in the cytoplasm or membrane due to hydrodynamic coupling between membrane, cortex and cytoplasm~\cite{Illukkumbura.etal2020}.
In addition, they can also induce cell-shape changes that lead to flows in the cytoplasm. 
For example, surface contraction waves during the maturation of starfish oocytes have recently been shown to induce such flows~\cite{Wigbers.etal2021, Klughammer.etal2018}.
Similarly, shape changes resulting from blebbing incidents coincide with intracellular flows~\cite{Charras.Paluch2008}.

\textit{The P\'eclet number} -- 
The relative impact of diffusion and flow on protein transport is quantified by the P\'eclet number ${\mathrm{Pe} = \xi {\cdot}v/D}$, where $v$ is the typical protein advection velocity and $\xi$ a characteristic length scale. 
Large values of the P\'eclet number correspond to protein transport that is dominated by flow rather than diffusion.
Hence, small proteins with large diffusion constants are less affected by flow than large proteins or protein assemblies.
In addition, the detailed chemical interactions of proteins with other biomolecules and cellular structures can affect the effective diffusivity and advection velocity~\cite{Agudo-Canalejo.etal2020}.
As for diffusive transport, the advection velocity -- and hence the P\'eclet number -- is a mesoscopic quantity that disregards the heterogeneous structure of the environment.
This approximation is justified since variations in the mobility coefficients within a given substrate are usually much smaller than the variations between different substrates, such as the cytoplasm and the membrane.
In general, a protein that diffuses in the cytoplasm is less affected by flows than it is when bound to the more viscous membrane.

\emph{Theory --}
The spatiotemporal transport of, and reactions between proteins are mathematically described by nonlinear partial differential equations (PDEs)~\cite{Frey.Brauns2020}. The protein dynamics in terms of their cytosolic (volume) concentrations $\mathbf{c}(\mathbf{r},t)$
and membrane (area) concentrations $\mathbf{m}(\pmb{\sigma},t)$ generally take the form of \textit{general transport equations} with flux and source terms
\begin{align}
    \partial_t \mathbf{c}(\mathbf{r},t)&=-\nabla\cdot\mathbf{J}_c+\mathbf{f}_\text{cyt}(\mathbf{c})\,,\label{eq:pdes-a} \\[1.2ex]
    \partial_t \mathbf{m}(\pmb{\sigma},t)&=-\nabla_{\mathcal{S}}^{}\cdot\mathbf{J}_m+\mathbf{f}_{\text{mem}}(\mathbf{m},\mathbf{c}|_{\mathcal{S}}) \,,
\label{eq:pdes-b}
\end{align}%
which represent a broad and general class of interesting dynamic systems far from thermodynamic equilibrium.
The divergence of the cytosolic and membrane fluxes $\,\mathbf{J}_{c/m}$ accounts for the (mass-conserving) spatial transport of proteins, and generally contains both diffusive and advective contributions. Here $\nabla_{\mathcal{S}}$ denotes the \emph{covariant derivative} 
for the \emph{curvilinear coordinates} ${\pmb{\sigma} \in \mathcal{S}}$ on the membrane surface $\mathcal{S}$.
The membrane is often considered as a static object for simplicity, however models can in general be extended to dynamic surfaces.
In particular, this requires to extend the dynamics by an explicit expression for the time evolution of the membrane geometry, $\mathcal S \to \mathcal S(t)$~\cite{Rangamani.etal2014,Wu.etal2018,Mietke.etal2018, Alimohamadi.Rangamani2018,Mietke.etal2019, Mahapatra.etal2021}.
The source terms $\mathbf{f_\text{cyt}}$ and $\mathbf{f_\text{mem}}$ result from the chemical reactions of the underlying protein networks, as discussed above.
Note that membrane-bound proteins not only react with each other, but membrane reactions also involve interactions with cytosolic proteins in close proximity to the membrane ($\mathbf{c}|_{\mathcal{S}}$).

The set of nonlinear PDEs (Eqs.~\eqref{eq:pdes-a} and~\eqref{eq:pdes-b}) is closed by \emph{reactive boundary conditions} at the membrane
\begin{equation}
   \left.\mathbf{J}_c\cdot \hat{\mathbf{n}}\right|_\mathcal{S}=\mathbf{g}(\mathbf{m},\mathbf{c}|_{\mathcal{S}}) \,,
\end{equation}
which ensures local mass conservation: cytosolic fluxes normal to the membrane ($\hat{\mathbf{n}}$ denotes the outward normal vector) must be balanced by reactive fluxes $\mathbf{g}(\mathbf{m},\mathbf{c}|_{\mathcal{S}})$ at the membrane~\cite{Frey.Brauns2020}.
An additional constraint for many models of protein pattern formation is the global conservation of protein mass, i.e., the assumption that no proteins are produced or degraded on the time scale of pattern formation.
This assumption is violated on longer time scales, where protein production and degradation processes -- in particular gene expression -- need to be taken into account~\cite{Goryachev.Leda2020}.

\subsection{Lateral instabilities and trigger waves}%
This set of general transport equations provides the theoretical framework for studying the spatiotemporal dynamics of protein patterns. 
The interested reader may consult recent lecture notes~\cite{Frey.Brauns2020} for an introduction to their analysis.
Here, to conclude our introduction to the basic principles of pattern formation, we briefly introduce two particularly interesting phenomena: pattern-forming instabilities and trigger waves.

A \textit{pattern-forming instability} arises when a spatially uniform steady state becomes unstable against spatially inhomogeneous perturbations (Fig.~\ref{fig:shape_size}d).
One example of such a pattern-forming instability is a mass-redistribution instability (see Supplementary Information), which amplifies spatial variations in protein number, thus leading to a protein concentration pattern~\cite{Halatek.Frey2018}.
The dynamics and length scale of these patterns on short time scales are determined by the growth rate and wavelength of the unstable modes, termed \emph{dispersion relation} (see Supplementary Information).
The growth rate of the unstable modes depends on the specific reaction kinetics and transport properties of the dynamics.
The wavelength of the fastest growing unstable mode determines the characteristic length scale of the initially growing pattern.
While the initial pattern is dominated by the dynamics of the unstable modes, the dynamics on longer timescales may be dominated by other processes, such as coarsening~\cite{Brauns.etal2021a} and non-linear interactions of the unstable modes far away from the linear regime.

In addition, nonlinear protein reaction kinetics can give rise to several reactive equilibria at the same total protein concentration, which is a necessary requirement for \emph{trigger waves}.
This phenomenon is best exemplified by systems that show bistability (see Supplementary Information)~\cite{Gelens.etal2014}.
In this case, the system can be at different reactive equilibria at different regions in the cell, giving rise to front-like protein activity patterns.
Such front-like patterns propagate with a finite velocity, whose magnitude and sign depend on the details of the reaction kinetics~\cite{Saarloos2003,Wigbers.etal2021}. 
This propagation is constrained by the limited abundance of proteins, which can result in localized wave fronts in cells~\cite{Mori.etal2008, Walther.etal2012, Rulands.etal2013}.
Moreover, unstable reactive equilibria can give rise to spatially homogeneous oscillations and traveling spiral waves~\cite{Halatek.Frey2018, Ferrell.etal2011, Paquin-Lefebvre.etal2020}.

The spatiotemporal properties of these patterns, such as the orientation of static patterns or the direction of propagating wave fronts, need to be controlled tightly by the cell.
This is achieved with the aid of guiding cues.
In the following, we will discuss the most prominent types of guiding cues observed to play a role in pattern formation processes in cells.

{
\scriptsize
\begin{figure*}[!t]
    \centering
    \includegraphics[width=0.85\textwidth]{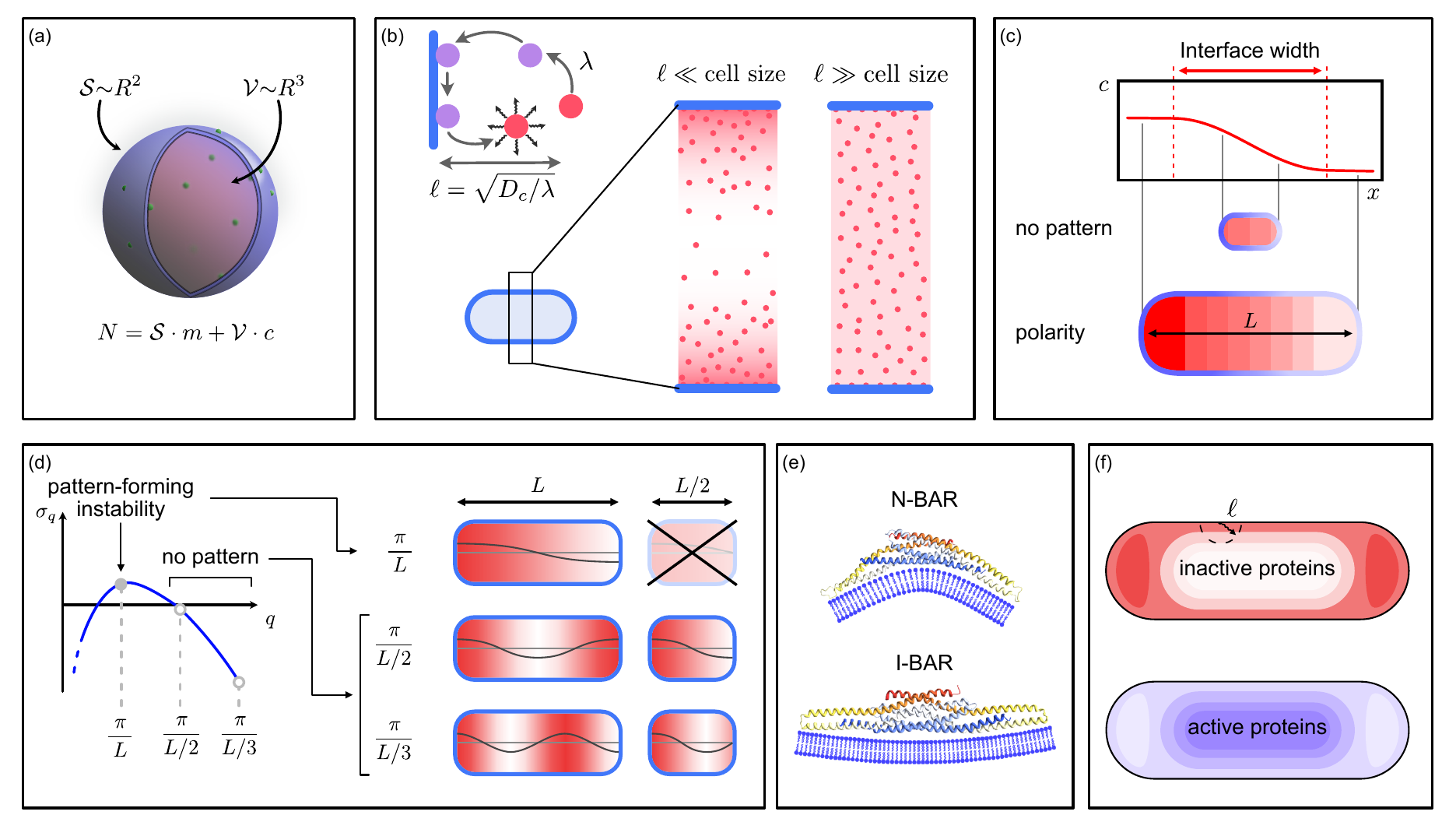}
    \caption{\textbf{Size and shape as guiding cues:}
    (a) Schematic illustration of protein distribution in the cytosol and on the membrane: the cell volume scales with cell size $R$ as  $R^3$, whereas the cell surface scales as $R^2$, implying that both membrane and cytosolic protein concentration change with cell size.
    (b) Left: cytosolic gradients can emerge when proteins undergo a `reacivation' step after detaching from the membrane. Inactive proteins (red) diffuse over a characteristic length scale $\ell$ before being reactivated (purple). Right: cytosolic gradients are established when the cell size is much larger than this characteristic length scale.
    (c) Cell size controls pattern formation: protein patterns cannot be established in cells smaller than the characteristic length scale of a pattern.
    (d) Only certain unstable modes with a wavelength limited by the cell size $L$ can be realised. In a cell of size $L/2$, no pattern-forming instability arises.
    (e) Proteins including BAR domains preferentially bind to similarly curved membranes.
    (f) Characteristic distribution of proteins with delayed reactivation in elongated cells. Inactive proteins are reactivated after diffusing over a characteristic length scale $\ell$. At the cell poles, this leads to the accumulation of inactive proteins, while they are diluted at the center of the cell. A complementary distribution of active proteins is established.}
    \label{fig:shape_size}
\end{figure*}
}
\section{Geometric guiding cues}%
On the largest scales, cells are characterized by their size and shape, which together confine protein transport and protein reaction kinetics. 
 
\subsection{Cell size controls protein patterns}%
Experimental studies show that, in addition to reaction and transport properties of the cell, also the cell size affects protein patterns. Examples include the transition from pole-to-pole oscillatory patterns to stripe patterns of MinD in filamentous \textit{E. coli} cells~\cite{Raskin.Boer1999, Zieske.Schwille2014}, and the observation that the PAR proteins in \textit{C.~elegans} fail to polarize in small cells~\cite{Hubatsch.etal2019}.

\textit{Bulk-boundary-ratio.--} On the time scale of pattern formation and dynamics, the total concentration of proteins remains constant.
As a consequence of these resource limitations, protein concentrations on the membrane and in the cytosol will in general depend on the ratio of membrane area to cell volume.
Moreover, the number and stability of reactive equilibria, as well as pattern-forming instabilities, are controlled by the total concentration of proteins (see Supplementary Information), and variations in cell size can therefore qualitatively affect protein patterns.
To understand the underlying idea, we assume for simplicity that the concentrations of cytosolic proteins $c$ and membrane-bound proteins $m$, respectively, are uniformly distributed.
The total number of proteins $N$ is then  given by ${N=\mathcal{S}\cdot m+\mathcal{V}\cdot c}$, where $\mathcal{S}$ and $\mathcal{V}$ denote the membrane (surface) area and the cytosolic (bulk) volume, respectively (Fig.~\ref{fig:shape_size}a).
Rewriting this mass-conservation relation in terms of the total protein density ${\rho=N/\mathcal{V}}$, one obtains ${\rho=\mathcal{S}/\mathcal{V}\cdot m+c}$.
Thus, the protein concentrations on the membrane and in the cytosol depend on the ratio of membrane to volume $\mathcal{S}/\mathcal{V}$;
for example, for a spherical cell with radius $R$, one finds ${\rho = 3\, m/R + c}$.

\textit{Cytosolic protein gradients.--}
Because the proteins of interest here are not permanently fixed to either the membrane or the cytosol, but circulate between these compartments due to various chemical processes such as membrane detachment, attachment, and recruitment, the cell membrane effectively acts both as a source and sink for cytosolic proteins.
These chemical reactions need to be balanced by diffusive fluxes in the cytosol, otherwise local mass conservation would be violated.
Hence, on these very general grounds, spatial gradients in the cytosolic protein density must be assumed~\cite{Frey.Brauns2020,Halatek.Frey2018}.
Strikingly, these gradients generally do not equilibrate over time, but are maintained by an interplay between diffusion and non-equilibrium reaction kinetics (see Supplementary Information).

Indeed, a good example is the case where proteins in the cytosol can have two different conformations, an \emph{inactive} and an \emph{active} state. 
Only proteins in the active state are able to bind to the membrane, and they typically undergo a conformational change to the inactive state upon detachment from the membrane (Fig.~\ref{fig:shape_size}b). 
In the cytosol, inactive proteins can switch back to the active state with a rate $\lambda$.
This reactivation step requires the consumption of energy and is a generic feature in NTPase or phosphorylation/dephosphorylation cycles~\cite{Bokoch.etal1994,Ubersax.FerrellJr2007,Irazoqui.etal2003}.
Since detached proteins cannot immediately bind to the membrane again, a protein concentration gradient may form in the cytosol~\cite{Kiekebusch.etal2012, Kiekebusch.Thanbichler2014}. 
The \emph{penetration depth} $\ell$ of this gradient depends on the cytosolic diffusion constant $D_c$ and the reactivation rate $\lambda$, and is given by ${\ell=\sqrt{D_c/\lambda}}$~\cite{Thalmeier.etal2016}.

If the cell size is much smaller than this penetration depth, the cytosolic protein concentration is effectively nearly homogeneous throughout the cell.
Conversely, if the cell is much larger than the penetration depth, protein gradients can be established in the cytosol~(Fig.~\ref{fig:shape_size}b).
The presence of such cytosolic gradients can fundamentally affect the formation of patterns on the membrane~\cite{Halatek.Frey2018, Frey.etal2018, Brauns.etal2021}.
This is well exemplified in the \textit{E.coli} Min system, which shows standing wave patterns \textit{in vivo}, but -- strikingly -- produces traveling and spiral wave patterns, among others, in reconstituted \textit{in vitro} assays with large bulk volume~\cite{Loose.etal2008, Zieske.Schwille2014, Ramm.etal2019}.

\textit{Finite size effects.--} In addition, cell size can affect pattern-forming instabilities.
A pattern-forming instability arises when a spatially uniform steady state is unstable against spatially inhomogeneous perturbations (Fig.~\ref{fig:shape_size}d).
Due to the finite size of the cell, only particular unstable modes can grow, where the largest possible wavelength is constrained by the lateral length of the cell.
Thus, while a reaction network can lead to a pattern-forming instability in large cells, it may result in a stable and spatially uniform steady state or a weak gradient in small cells (Fig.~\ref{fig:shape_size}c,d).
Indeed, this has been observed for the polarity pattern of PAR proteins in \textit{C.~elegans}  (Fig.~\ref{fig:shape_size}c)~\cite{Hubatsch.etal2019}.
Similarly, cell size may not only limit the existence of a pattern, but also the type of protein pattern that can be established.

\subsection{Cell shape and curvature sensing}
\label{sec:shape}
For a wide range of cells, from bacteria~\cite{Wu.etal2016, Raskin.Boer1999, Varma.etal2008} to migrating fibroblasts~\cite{Begemann.etal2019} to unicellular eukaryotes~\cite{Mishra.etal2012} and large zygotes~\cite{Klinkert.etal2019}, cell shape and local membrane curvature serve as important guiding cues for protein attachment to the membrane.
The mechanisms underlying such curvature detection are based on the interaction of proteins with the membrane, in particular its membrane binding affinity (\emph{curvature-sensing proteins}), and the probability that a protein will make contact with the membrane (\emph{collective curvature sensing}).
Both factors can be affected by cell shape (membrane curvature).

\subsubsection{Curvature-sensing proteins}%
One prominent set of proteins that can individually sense membrane curvature are proteins containing a curved BAR domain\footnote{BAR domain -- A curved protein domain that binds to curved membranes, named after three proteins that contain this domain: Bin, Amphiphysin, and Rvs.}~\cite{Mim.Unger2012, Simunovic.etal2015, Peter.etal2004, Peleg.etal2011}.
These proteins preferentially bind to membrane regions that have a curvature comparable to that of the BAR domain itself (Fig.~\ref{fig:shape_size}e). 
For example, during persistent cell motion, the curvature-sensitive protein BAIAP2, which contains such a BAR domain, accumulates at curved membrane patches at the cell front, inducing the formation of lamellipodia~\cite{Begemann.etal2019}.
Since BAR domains have a length of about 20$\,$nm, the sensitivity of individual proteins to weakly curved surfaces is limited~\cite{Qualmann.etal2011,Simunovic.etal2015}.
However, membrane curvature can facilitate the oligomerization of proteins into extended curved structures, which are capable of sensing membrane curvature on length scales larger than that of the individual protein~\cite{Feddersen.etal2021}.
Other important examples for such joint curvature sensing are dynamin, which forms helical collars around the thin neck during budding in yeast~\cite{Faelber.etal2012, Shlomovitz.etal2011}, and MreB, which assembles into filaments that orient along the highest membrane curvature~\cite{Hussain.etal2018, Wong.etal2019}.

Furthermore, some proteins recognize membrane curvature via defects in membrane structure.
This mechanism is well exemplified by proteins with so-called \emph{ALPS motifs}.
ALPS motifs do not have a defined structure in solution, but insert into lipid bilayers by folding into an $\alpha$-helix\footnote{$\alpha$-helix -- Prevalent helical-like protein structure, which is highly stable due to hydrogen bonds.}.
It has been shown that ALPS motifs bind preferably to regions with low lipid packing density~\cite{Antonny2011}.
Such low-density packing can arise from membrane curvature, where one sheet of the lipid bilayer is stretched compared to a flat membrane.
In experiments, ALPS motifs were found to bind strongly to liposomes with sufficiently strong positive curvature (${R< 50~\mathrm{nm}}$), and to weakly curved liposomes with a high concentration of conically shaped lipids~\cite{Antonny2011}.
Thus, curvature-dependent binding affinity can lead to predominant accumulation of proteins at curved membrane regions.

It has been reported that proteins that sense curvature can also deform the membrane:
The helical structure of dynamin oligomers induces membrane curvature during scission of the yeast bud~\cite{Faelber.etal2012, Ferguson.Camilli2012, Roux.etal2010}.
Proteins with BAR domains play a curvature-sensing role at low concentrations, but stabilize membrane curvature at high protein concentrations~\cite{Simunovic.etal2015, Mim.Unger2012}.
Such a dual role can lead to a positive feedback loop, when a slightly curved membrane leads to the accumulation of curvature-sensitive proteins. 
These proteins, in turn, deform the membrane, leading to a further increase in the binding affinity. 
This has been proposed as a general mechanochemical mechanism for protein recruitment~\cite{Goychuk.Frey2019}.
However, the formualation of a mechanistic theory for such curvature-regulating feedback loops remains an open and highly interesting challenge to this day.

\subsubsection{Collective curvature sensing}%
It has recently been shown that the distribution of proteins on the membrane and in the cytosol can depend on the cell geometry, even when the binding affinity of proteins is independent of membrane curvature~\cite{Thalmeier.etal2016, Gessele.etal2020, Eroume.etal2021}.
The underlying mechanism is based on the aforementioned cytosolic gradients of proteins that switch between an inactive and an active state in the cytosol.
As the required reactivation step is a non-equilibrium process that consumes energy, these gradients are maintained by a constant cycling of such proteins between the membrane and the cytosol, and therefore do not equilibrate by cytosolic diffusion.
Since cytosolic gradients from opposing membrane points overlap at curved regions, one generally expects accumulation of inactive proteins in regions of high curvature (e.g., near the cell poles of elongated cells, including the rod-shaped \textit{E.~coli}~\cite{Thalmeier.etal2016}, the \textit{C.~elegans} zygote~\cite{Gessele.etal2020}, and \textit{Bacillus subtilis}~\cite{Feddersen.etal2021}) and a corresponding depletion of active proteins (Fig.~\ref{fig:shape_size}f).
Moreover, the effect of such a cytosolic gradient on the protein distributions in curved geometries depends in particular on the characteristic length $\ell$ of the cytosolic gradient relative to the local membrane curvature~\cite{Thalmeier.etal2016, Gessele.etal2020}.

While this explains where proteins are most likely to encounter the membrane, its effect on the ensuing protein pattern depends on the protein reaction kinetics.
For proteins that exhibit a simple attachment-detachment dynamics with the membrane, the increased encounter probability leads directly to an increase in protein concentration at the poles, which is further enhanced if the protein autocatalytically promotes its own binding~\cite{Thalmeier.etal2016}.
In contrast, if two proteins mutually inhibit each others binding, an increased encounter probability leads to the formation of an interface between two protein domains on the membrane~\cite{Gessele.etal2020}.
\begin{figure*}[t]
    \centering
    \includegraphics[width=0.85\textwidth]{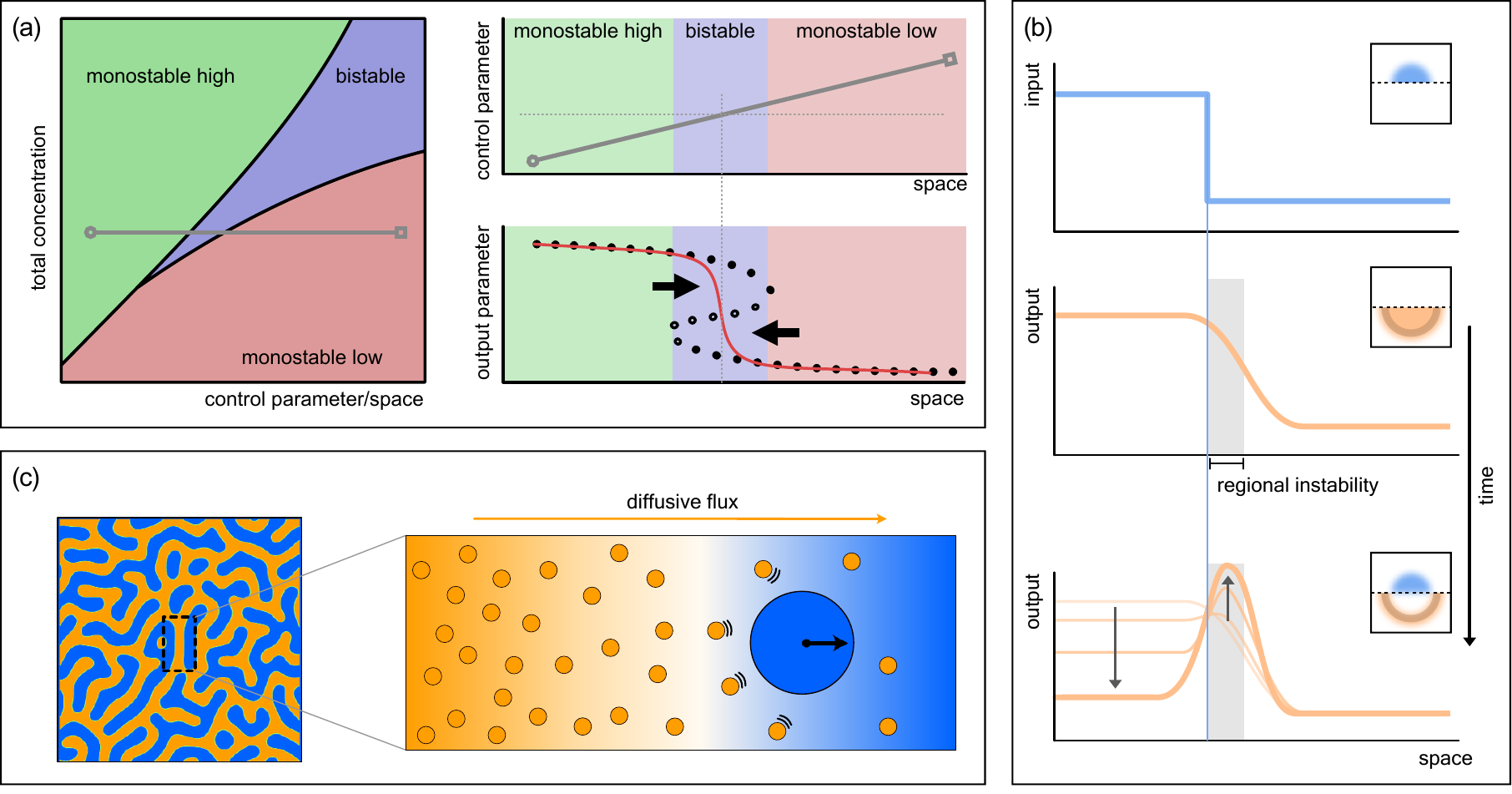}
    \caption{\textbf{Principles of biochemical pattern guidance:}
    (a) Left: Characteristic bifurcation diagram for pattern-forming systems.
    For reaction kinetics where the concentration of the input protein is a control parameter, a spatially varying input protein concentration can serve as a map between space and varying reaction kinetics.
    Top right: an input protein concentration gradient corresponds to a cutline through the bifurcation diagram (gray line) laid out in space, which divides the cell into regions of distinct stability.
    Bottom right: for a system where the input concentration gradient connects two monostable regions via a bistable region, the resulting front pattern (red line) is pinned to a threshold concentration value of the input concentration. Fixed points of the protein reaction kinetics are indicated by filled (stable) and open (unstable) circles.
    (b) Edge detection: An input pattern (blue) spatially alters the reaction kinetics of the output protein, resulting in a regional instability of the output protein close to the input edge (gray filled area). 
    This leads to a peak pattern of the output protein concentration (orange) that marks the position of the input edge.
    Insets show a possible realization of this edge-sensing process, leading to a ring around a template patch.
    The plots depict the concentration profiles along the black cutline.
    (c) Diffusiophoresis:
    Diffusive fluxes of pattern-forming proteins (carrier particles, shown in orange) are established at pattern interfaces.
    Carrier particles transport cargo particles (blue) via frictional interactions, resulting in a complementary pattern of cargo particles~\cite{Ramm.etal2021}.
    }
    \label{fig:biochemical}
\end{figure*}
\section{Biochemical guiding cues}%
For spatially homogeneous systems, several theoretical and experimental studies have identified biochemical circuits that are able to perform logic operations~\cite{Alon2007}, generate pulses~\cite{Basu.etal2004,Ishihara.etal2005}, act as noise-reduction filters~\cite{Barkai.Leibler1997}, or process biochemical signals in other ways~\cite{Tyson.Novak2010,Benenson2012,Alon2019, Bray1995,Purvis.Lahav2013}.
Here the information from an input signal -- typically encoded in the concentration of a protein -- is processed and an output signal is generated.

In general, however, protein concentrations tend to be spatially inhomogeneous, so that a locally varying input can lead to a locally varying output protein concentration in the cell.
In this way, an input pattern can serve as a template or \textit{biochemical guiding cue} for the formation of an output protein pattern.
Such biochemical guidance has been observed in many biological processes and over widely varying scales, ranging from tissue development~\cite{Gregor.etal2007,Strigini.Cohen2000} to the positioning of the cell-division site~\cite{Kiekebusch.Thanbichler2014, Chiou.etal2016, Martin.Berthelot-Grosjean2009, Schumacher.etal2017, Ramm.etal2019, Lutkenhaus2007, Tong.etal2007}.
In all these cases, the input patterns encode positional information, as each concentration marks a specific location or region in space~\cite{Wolpert1969}.
In fact, there are several known instances in which protein patterns (input) control the formation of other patterns (output)~\cite{Griffin.etal2011, Rodriguez.etal2017, Magliozzi.etal2020, Walker.etal2020, Kiekebusch.Thanbichler2014}.
However, the physical mechanisms responsible for the processing of the positional information encoded in patterns, and the generation of a qualitatively different output pattern (e.g., gradient vs.~step profile) are still largely unclear.

Such input/output relations are found, for example, in the polarity mechanism of budding yeast.
Here, several so-called \emph{landmark proteins} mark specific locations in the cell, such as the previous bud site.
These landmark proteins (input) alter the kinetics of nucleotide exchange in the polarity factor Cdc42 (output), and thus contribute to the control of cell polarity in a symmetry-breaking manner~\cite{Meitinger.etal2014, Bi.Park2012}.
Another example is provided by the midcell localization machinery of \textit{Caulobacter crescentus}.
In these elongated cells, ParB-\textit{parS} (input) complexes localized to the cell poles stimulate the ATP-dependent dimerization of MipZ (output), which results in the formation of a bipolar gradient of MipZ dimers with a minimum at midcell~\cite{Kiekebusch.etal2012}.
MipZ, in turn, inhibits the polymerization of FtsZ, which is a central component of the cell-division machinery.
Thus, the bipolar MipZ gradient also acts as an input for the control and positioning of FtsZ (output) to midcell~\cite{Thanbichler.Shapiro2006}.
Such a hierarchy of pattern control through multiple stages of protein interaction is a common feature of many biochemical guidance mechanisms~\cite{Wigbers.etal2021, Chang.FerrellJr2013, Martin.Berthelot-Grosjean2009, Bi.Park2012}.

In the following, we discuss some recent advances in this area, focusing on systems in which the concentration profile of an (input) protein is able to control the reaction kinetics of another (output) protein, such that one or more reaction rates become spatially inhomogeneous.
This can result in an output protein pattern that is qualitatively different from the input pattern, which has been termed \textit{spatial network computations}~\cite{Kinkhabwala.Bastiaens2010}.

\subsection{Spatially varying reaction kinetics}%
Since protein reaction kinetics can depend on the concentration of other proteins, a spatially varying input protein concentration can lead to locally varying reactive equilibria of the output protein.
In particular, not only can the protein concentration at each local reactive equilibrium be altered; also, the number and the stability of these equilibria can change in response to a varying input concentration (see Supplementary Information).
Heuristically, this means that space itself serves as a control parameter\footnote{Control parameter -- A parameter that alters the qualitative dynamics when it is changed, also referred to as a \emph{bifurcation parameter} in nonlinear dynamics.}
for the protein reaction kinetics.
Hence, the input protein pattern encodes positional information.

The dynamics of the output protein depend crucially on its explicit biochemical interactions with the input proteins.
For example, for a particular interaction between proteins, this can lead to bistability of the output protein over a limited range of input concentrations, as observed in starfish oocytes~\cite{Wigbers.etal2021}.
Due to the correspondence between input protein concentration and space, such a bistable parameter range maps to a region in space where the output protein reactions are bistable, which we refer to as \textit{regional bistability}.
In a similar way, a protein pattern can cause a pattern-forming instability in a specific spatial region, which has been termed \textit{regional instability}\cite{Brauns.etal2020,Wigbers.etal2020a}.
Thus, an input pattern can lead to a qualitatively different spatial concentration profile of the output protein, where the explicit output pattern strongly depends on the reaction kinetics (Fig.~\ref{fig:biochemical}a).
This fundamental property of protein interactions is likely to represent the mechanism that underlies many of the biochemically guided pattern-forming systems observed in experiments~\cite{Bi.Park2012, Magliozzi.etal2020, Meitinger.etal2014, Thanbichler.Shapiro2006, Walker.etal2020}.

\subsection{Wave localization by protein gradients}%
Biochemical trigger waves, consisting of a traveling front or pulse of biomolecule concentration, are a common means of long-ranged signal transmission in cells~\cite{Gelens.etal2014}.
Prominent examples of such waves include calcium waves~\cite{Falcke2007}, the propagation of mitosis\footnote{Mitosis -- Stage of the cell cycle during which chromosomes are segregated into the two daughter cells.}~\cite{Chang.FerrellJr2013} and apoptosis\footnote{Apoptosis -- Cellular process leading to actively induced cell death.}~\cite{Cheng.Ferrell2018} in \textit{Xenopus} eggs, actin polymerization waves in \textit{Dictyostelium}~\cite{Bretschneider.etal2004} and neutrophils~\cite{Houk.etal2012}, as well as intracellular signaling~\cite{Bezeljak.etal2020}.
A key component of models for trigger waves, such as the FitzHugh-Nagumo model~\cite{FitzHugh1961}, are bistable reaction kinetics (see Supplementary Information).
These bistable reaction kinetics, in addition to resulting in information transmission, allow trigger waves to serve as a readout for positional information encoded in other protein patterns.

To illustrate how spatially varying reactive equilibria allow proteins to read out this positional information, we now discuss how a protein gradient can lead to the localization of such a trigger wave, in particular a bistable front, to a specific position in the cell.
We first consider a system with homogeneous bistable reaction kinetics forming a front pattern (see Supplementary Information).
This front can propagate through the system at a speed and direction that depends on, among other factors, the concentration of the input protein~\cite{Gelens.etal2014,Mikhailov1994}.

In the presence of an input pattern, the reaction kinetics are no longer homogeneous, so that a regional bistability can emerge.
Since the front only propagates in a bistable parameter range, propagation is constrained to this regional bistability.
In particular, since the direction of propagation depends on the input concentration, the front is pinned at a threshold input concentration (Fig.~\ref{fig:biochemical}a)~\cite{Gelens.etal2014}.
Due to the correspondence between input concentration and space, this means that the front is localized to a specific position within the regional bistability.
Thus, the position of the front interface marks the location of the input threshold concentration, allowing the positional information encoded in the input pattern to be read out.
Such a threshold-sensing mechanism has been proposed to play a role in the propagation of surface contraction waves during meiosis\footnote{Meiosis -- A type of cell division process that generates daughter cells that contain half as many chromosomes as the parent cell.} in starfish oocytes~\cite{Wigbers.etal2021} and during chemotaxis\footnote{Chemotaxis -- Directed locomotion of cells along chemical gradients.} in eukaryotes~\cite{Beta.etal2008}.

\subsection{Edge-sensing and ring formation}%
Proteins also have been found to localize at the edges of spatial domains that exhibit a high concentration of other proteins or macromolecules. 
For example, during cellular wound healing, the Rho-GTPase Cdc42 and an associated GTPase regulator, Abr, accumulate locally to form two concentric rings~\cite{Vaughan.etal2011}. 
Experimental evidence suggests that this structure is hierarchically organized, with the outer Cdc42 ring being dependent on the presence of an inner Abr zone.
While it is not particularly surprising that a given spatial protein profile serves as a template for creating another protein profile with a similar shape, it is quite interesting that the downstream profile assumes a qualitatively different shape, with a peak localized right at the edge of the upstream profile (inner Abr ring, see insets in Fig.~\ref{fig:biochemical}b). 
To account for such edge-sensing, a regional instability has been suggested~\cite{Wigbers.etal2020a,Agudo-Canalejo.etal2020}.
Here, the step-like Abr profile, acting as an input protein pattern, defines two spatial domains with qualitatively different reaction kinetics for Cdc42, which takes the role of the output protein. 
It was shown that the outer domain may effectively act as a stimulus that induces a lateral mass-redistribution instability in the inner domain, which leads to a concentration peak of the output protein at the template edge (Fig.~\ref{fig:biochemical}b).
Moreover, the formation of this output concentration ring can be controlled by both the magnitude of the input pattern step and the total amount of output protein.
Thus, edge sensing is enabled by a regional mass-redistribution instability in a downstream protein pattern, which is itself triggered by an upstream protein pattern that acts as a step-like template.

Beyond the specific example discussed above, there are other biologically highly relevant processes that involve edge sensing.
As in the case of wound healing, a ring of Rho forms around a patch of high Cdc42 concentration prior to polar body emission in \textit{Xenopus} oocytes~\cite{Zhang.etal2008}.
Another biological process in which protein templates appear to play an essential role is that of macropinocytosis, a form of endocytosis\footnote{Endocytosis -- Cellular process that enables the uptake of biomolecules into the interior of the cell.} associated with cell surface ruffling. 
Here, actin-recruiting proteins colocalize to high-density patches of PIP3 (a charged phospholipid) and a Ras-GTPase, forming a ring around the edge of the PIP3 domain, which in turn leads to the assembly of a conctractile actomyosin ring~\cite{Veltman.etal2016}.
This whole process is invariably linked to the presence of PIP3 and Ras patches, suggesting that these biomolecules serve as a biochemical guiding cue for the actin-recruiting proteins.
The specific physical mechanisms responsible for each of these edge-sensing processes have not yet been uncovered.

\subsection{Tracking of moving patterns}
In addition to varying in space, the input protein concentration can vary in time at a fixed location in the cell.
Temporal changes of the input concentration can lead to sudden changes of the reactive equilibrium which, in turn, results in transient dynamics of the output concentration before the new reactive equilibrium is established -- a phenomenon referred to as \textit{excitability} in the field of nonlinear dynamics~\cite{Cross.Greenside2009,Mikhailov1994}.
Such transient dynamics can mark the position of local changes in the input concentration.
For example, in the case of a traveling front pattern, the input concentration changes in time at a fixed position as the front passes by.
Due to the transient output dynamics, this can lead to a traveling output concentration peak that closely follows the moving front.
This has been observed in starfish oocytes, where a traveling front pattern leads to a moving concentration peak which, is ultimately responsible for the surface contraction waves observed during meiosis~\cite{Bement.etal2015,Bischof.etal2017,Wigbers.etal2021}.
Similar observations have been made \textit{in vitro} for an artificial cortex based on frog egg extracts~\cite{Landino.etal2021}.

\subsection{Phoretic transport}
A more intricate mechanism by which spatiotemporal protein patterns could serve as cues for the development of subsequent protein patterns are various types of phoretic transport processes.
These are, in general, the result of an external field gradient acting on the protein~\cite{Marbach.Bocquet2019, Anderson1989}.
Examples include concentration gradients of carrier particles (\emph{diffusiophoresis})~\cite{Derjaguin.etal1993, Ramm.etal2021}, chemical potential gradients (\emph{chemophoresis})~\cite{Vecchiarelli.etal2014, Sugawara.Kaneko2011}, electric potential gradients (\emph{electrophoresis})~\cite{Allen.etal2013}, or temperature gradients (\emph{thermophoresis})~\cite{Iacopini.Piazza2003}, along which cargo can be transported.
Thus, cargo particles can form a pattern guided by such gradients~\cite{Marbach.Bocquet2019}.
Notably, in phoretic transport mechanisms, energy is consumed to maintain the gradient, resulting in a flux of cargo particles.
This is substantially different from other transport mechanisms such as active transport, where energy is consumed to fuel molecular motors that move cargo particles.

In the field of phoretic transport, research has long been focused on colloidal particles~\cite{Marbach.Bocquet2019, Derjaguin.etal1993, Anderson1989, Palacci.etal2011}.
Experimental evidence for phoretic transport in biological systems related to protein organisation and pattern formation has only recently been discovered~\cite{Ramm.etal2021, Vecchiarelli.etal2014}.
For example, \textit{in-vitro} experiments have shown that diffusiophoresis can result in the spatial organization of DNA origami nanostructures in a concentration gradient of MinD~\cite{Ramm.etal2021}.
Here, the Min proteins self-organize into a stationary pattern~\cite{Glock.etal2019}, resulting in diffusive fluxes at the domain edges (c.f. Fig.~\ref{fig:biochemical}c).
These diffusive fluxes are transferred to the DNA nanostructures via friction, leading to diffusiophoretic transport of the latter along the Min gradients.
Thus, the movement of the DNA nanostructures mimics the movement of the Min proteins, resulting in the formation of an anti-correlated pattern of the DNA nanostructures.
Such diffusiophretic transport has been suggested to play an important role for the distribution of large particles in cells in general~\cite{Sear2019}.

In the context of plasmid segregation, chemophoresis has been suggested to drive the movement of plasmids on the nucleoid~\cite{Vecchiarelli.etal2014}.
Here, ParA proteins on the nucleoid surface are thought to bind to large cargo, such as plasmids.
Upon unbinding, ParA proteins are released from the nucleoid, resulting in a local depletion of ParA at the position of the cargo.
The ParA concentration gradient at the edge of this depletion zone creates a chemical potential gradient for the cargo, which tends to bind more strongly at regions of high ParA concentration.
Thus, the cargo moves along the chemical potential gradient away from the depletion zone~\cite{Vecchiarelli.etal2014, Sugawara.Kaneko2011}.
This chemophoretic movement is suggested to be sufficient to ensure a balanced distribution of plasmids on the nucleoid~\cite{Vecchiarelli.etal2014}.

\section{Mechanical guiding cues}%
\begin{figure*}
    \centering
    \includegraphics[width=0.85\textwidth]{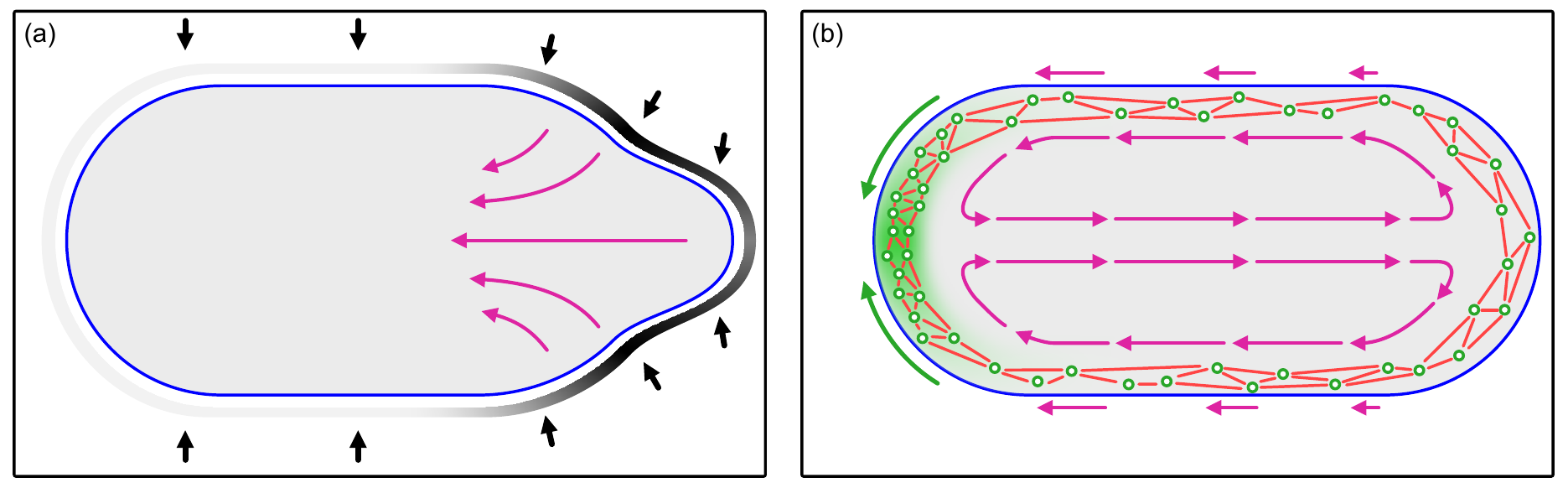}
    \caption{\textbf{Principles of mechanical guidance by flow generation:} Stress gradients result in flows.
    (a) Heterogeneous cell deformations, as indicated by the grayscale outline, lead to pressure gradients in the cytosol, which in turn induce cytosolic flows towards regions of low pressure.
    (b) Heterogeneous actomyosin activity (green gradient; actin filaments shown in red, myosin shown as green circles), as observed in \textit{C.~elegans} zygotes~\cite{Gubieda.etal2020}, leads to polarized contractions of the actomyosin cortex and a flow of the entire cortex towards regions of high actomyosin activity. Hydrodynamic coupling results in cytosolic flows.}
    \label{fig:mechanics}
\end{figure*}
In addition to biochemical guiding and guidance by cell size and shape, also the mechanical properties of a cell can affect protein pattern formation by altering the transport and reaction kinetics of proteins.

Flows generally arise from stress gradients.
In cells, such gradients can be generated via shape deformations (Fig.~\ref{fig:mechanics}a).
For example, recent work has demonstrated the generation of flows in the cytoplasm due to shape deformations in starfish oocytes~\cite{Klughammer.etal2018}.
In these cells, a \emph{surface contraction wave} travels across the membrane from the animal to the vegetal pole,
which locally increases the pressure in the cytosol, and results in cytoplasmic flows along the oocyte's animal-vegetal axis\footnote{Animal-vegetal axis -- Symmetry axis in oocytes, along which the developmental activity varies, separating the cell into two distinct poles.}.
Similar observations have been made for \textit{Drosophila} embryos, where apical constrictions instead of surface contraction waves lead to cytoplasmic flows~\cite{Streichan.etal2018}, and in \textit{Drosophila} neuroblasts where cortical contractions induce flows in the cortex~\cite{Oon.Prehoda2019}.

Next to deforming the cell shape, contractions of the actomyosin cortex can also lead to cortical flows, either as a consequence of spatially inhomogeneous actomyosin activity~\cite{Gross.etal2019} or anisotropic cortical tension~\cite{Mayer.etal2010}  (Fig.~\ref{fig:mechanics}b).
For example, cortical flows in \textit{C.~elegans} zygotes prior to PAR polarization arise due to nonuniform actomyosin activity~\cite{Gross.etal2019}.
Through hydrodynamic coupling, such flows may also induce cytoplasmic flows~\cite{Gubieda.etal2020,Illukkumbura.etal2020}.

How are protein patterns controlled by mechanical guiding cues?
It has been suggested that a combination of pattern guidance by cortical flows and biochemical interactions may be ultimately responsible for the polarization mechanism in \textit{C.~elegans} zygotes~\cite{Gross.etal2019}.
Prior to polarization, a mechanical inhomogeneity in the cell cortex, induced by the symmetry-breaking introduction of a centrosome into the zygote, causes the cell cortex to contract asymmetrically.
Here, the reduced actomyosin contractility at the posterior pole leads to anterior-directed cortical flow.
Once symmetry is broken, the cortical flows and the associated anterior-directed cytoplasmic flows lead to a redistribution of PAR proteins, which in turn control and maintain the asymmetric actomyosin contractility of the cortex, thereby giving rise to a self-regulating polarization mechanism.
These observations underline the key role of mechanical guiding cues in the process of protein pattern formation.

\section{Upcoming challenges}%
In this review, we have focused on guidance mechanisms in model biological organisms that have been studied experimentally, and for which theoretical models exist.
However, a much larger number of cellular processes rely on guiding cues and whose underlying biophysical mechanisms are still unknown.
To conclude this review, we outline some promising recent developments in the field of protein pattern formation that build upon the recognition of the important role of guiding cues.

\subsection{Robustness against guiding cues}%
Guiding cues can vary over time, as evidenced by cell size and shape, which change throughout the cell cycle.
Moreover, these changes can affect the process of protein pattern formation in quite different ways:
Protein patterns can either adapt to the changing guiding cues as discussed in this review, or they can be impervious to variations in geometric, mechanical, and biochemical factors.
Pattern-forming mechanisms that are robust to changes in cell geometry or mechanics have recently been identified in various systems~\cite{Wigbers.etal2021, Alon2019}, but a general understanding of robustness in pattern formation is still lacking.
Future research on pattern formation mechanisms in living cells will reveal whether there are more examples where the formation of protein patterns adapts to be robust to the effects of cell mechanics and geometry.

\subsection{Mechanochemical feedback loops}%
We discussed above how protein patterns can flexibly adjust to changes in the physical properties of cells.
However, proteins can also actively modify the mechanical properties of the cell, resulting in a feedback loop between cell mechanics and protein patterns.
Various theoretical studies showed that the coupling to cell mechanics in such mechanochemical feedback loops can lead to the formation of protein patterns~\cite{Mietke.etal2018, Brinkmann.etal2018, Miller.etal2018, Cagnetta.etal2018, Gov2018, Tozzi.etal2019, Bois.etal2011}.
For example, coupling of a contractility-regulating chemical agent to an active fluid surface can result in shape deformations of axisymmetric surfaces, accompanied by polarization of the chemical agent~\cite{Mietke.etal2019}.
This phenomenon shows similarities to the aforementioned self-reinforcing polarity mechanism of \textit{C.~elegans}, where cortical flows are created by asymmetric actomyosin activity~\cite{Mayer.etal2010}.
In addition, a recent experimental study showed that the spatiotemporal patterning of the Min protein system can induce substantial shape deformations in GUVs\footnote{GUV -- Giant unilamellar vesicle, an artificial spherical chamber bounded by a lipid bilayer that mimics the membrane of cells.}~\cite{Christ.etal2020, Litschel.etal2018}.
This observation suggests a generic interplay between reaction-diffusion dynamics and membrane mechanics.
We hypothesize that membrane properties, such as spontaneous curvature, may influence the kinetics of protein binding, and vice versa~\cite{Mahapatra.etal2021, Rangamani.etal2014, Goychuk.Frey2019}.
In combination with the hydrodynamic coupling of the cell membrane to the cortex and the cytosol, this can lead to a mutual feedback between the dynamics of protein patterns and cell shape.

A theoretical characterization of this two-way coupling between biochemical processes and cell mechanics is a promising avenue for future research~\cite{Wigbers.etal2020}.
Since such mechanochemical models need to account for protein reaction–diffusion dynamics as well as a dynamically varying three-dimensional cell shape, they are challenging to study both analytically and numerically~
\cite{Brinkmann.etal2018,Tozzi.etal2019, Hoerning.Shibata2019, Salbreux.Juelicher2017}.
In future research, it will be important to further develop methods and, in particular, biologically realistic three-dimensional models, such that they can be compared to quantitative experimental data and contribute to the interpretation of experimental results in mechanochemical model systems.

Mechanochemical feedback loops are a special case of a general phenomenon that can be observed in many pattern-forming systems:
may patterns in cells are not the result of a single guiding cue, but are the products of multiple interacting cues and processes~\cite{Maree.etal2012, Chiou.etal2016, Gubieda.etal2020, Minc.etal2009, Rogez.etal2019, Scott.etal2021}.
However, it is often difficult to separate all the processes involved in the robust formation of functional protein patterns in living cells, as the example of \textit{C.~elegans} polarisation shows~\cite{Gubieda.etal2020, Gessele.etal2020, Klinkert.etal2019, Hubatsch.etal2019}.
Recognizing and incorporating such interacting processes into the theoretical analysis of pattern-forming systems will therefore be a major task for future research on pattern formation.

\subsection{Perspectives for pattern guidance}%
At the conceptual level, we currently face three main challenges in the context of understanding the biophysical basis of pattern guidance.
These relate to (i)~progress in the study of fundamental aspects of processes in living systems far from thermal equilibrium, (ii)~finding the right level of simplification for a given complex biological system, and (iii)~improving both computational and experimental tools.
In the long term, meeting these challenges will be vital to advancing our knowledge of pattern guidance, pattern formation, and information processing in biology in general.

\subsubsection{New frontiers in non-equilibrium physics}%
Several interesting physics questions arise from the biological model systems we have discussed in this review.
A central issue concerns how the dynamics of pattern-forming systems are mechanistically controlled by spatial and temporal gradients. 
These gradients lead to a variety of fascinating phenomena including information processing~\cite{Gregor.etal2007}, templating~\cite{Wigbers.etal2020a}, and hierarchies of different patterns~\cite{Wigbers.etal2021}.
Since these gradients can form for different physical quantities 
they can influence the formation of patterns in many ways.
Among others, we have discussed spatially varying reaction kinetics which can lead to the localization of trigger waves in bistable media. 
But any gradient in an intensive thermodynamic variable, such as a chemical potential, can give rise to corresponding particle currents, as described by the laws of non-equilibrium thermodynamics~\cite{Groot.Mazur2013}.
Transport properties are also strongly influenced by spatial variations in kinetic coefficients such as diffusion constants. 
These processes lead to additional advection currents which we have not addressed in this review. 
Moreover, due to dynamic feedback between these particle currents and protein patterns, the gradients themselves may become part of the dynamics rather than acting solely as external guiding cues.
This greatly expands the possibilities for future theoretical and experimental research on this topic.

\subsubsection{Levels of biological complexity}%
Another crucial and actually quite general challenge is how to deal with the different levels of complexity in biological systems.
For example, the full extents of interaction networks of proteins are generally unknown, and it is often unclear whether integrating all possible interaction pathways into a theoretical model is actually necessary to explain a particular phenomenon~\cite{Motegi.Seydoux2013, Gutenkunst.etal2007}.
Even in cases where networks are fully characterized, the information flow through the reaction network can be difficult to understand.
Methods to analyse such information flows have been developed for well-mixed reaction systems, such as the \emph{modular response analysis}~\cite{Bruggeman.etal2002}.
For spatially extended systems, where information is stored and processed by patterns, such methods have yet to be developed.

In addition, temporal regulatory mechanisms, such as cell-cycle-induced gene regulation, are often excluded from models of pattern forming system, even though the relevance of such regulatory mechanisms for pattern formation is not fully understood yet~\cite{Ishihara.etal2005, Reich.etal2019}.
Where such mechanisms are in place, global mass conservation -- which is a cornerstone in many models of protein pattern formation -- does not apply anymore, opening an avenue to additional concepts for pattern formation~\cite{Brauns.etal2021a}.

Avoiding the overfitting of models, and separating important components of interaction networks from irrelevant interactions (on the time scale of interest), are both difficult to achieve, and this presents major difficulties for theory and mathematical modeling.
Ultimately, theoretical frameworks need to be developed that allow for a systematic coarse-graining that shows how the manifold components of a biological system can be reduced to its core elements.
Such reductionism, at least for someone trained in physics, is the silver bullet to determining fundamental principles and improving our understanding.

\subsubsection{Finding the right level of geometric representation}%
Similarly, the question of how theory should deal with the dimensionality and geometric form of biological systems needs careful consideration.
For example, reducing the dimension of a specific system, e.g., to simplified one-dimensional models, may help to obtain an analytically more accessible representation.
While such a simplification can be useful for gaining insight into the underlying dynamics and for guiding experiments, it may also obscure important aspects of pattern guidance.
As pointed out in this review, certain phenomena, such as curvature sensing, only occur in realistic geometries and would therefore be erased in simplified one-dimensional models~\cite{Thalmeier.etal2016, Gessele.etal2020}.
In essence, the complexity of biological systems must be reduced in order to understand them better.
However, the challenge for future models is to find the appropriate level of simplification without loss of crucial features.

\subsubsection{How to face the challenge of multiphysics problems}%
In addition, many experimental results indicate that pattern formation, and pattern guidance in particular, are the result of a tight interplay between biochemical interactions, hydrodynamics of cellular substrates, and membrane mechanics~\cite{Gross.etal2019, Begemann.etal2019, Mayer.etal2010, Tan.etal2018}.
While numerous theoretical advances have been made in each of these areas (e.g., reaction-diffusion dynamics and non-equilibrium physics), there is so far no unified theoretical and computational approach that would allow a thorough analysis of such \emph{multiphysics} problems.
Therefore, in order to gain a deeper understanding of pattern guidance in realistic biological systems, a comprehensive theoretical framework that allows the study of the interplay between these different fields of physics must be developed.

\subsubsection{Improving experimental and computational methods}%
Another roadblock that impedes progress is the limited availability of experimental, analytical and computational tools.
On the experimental side, the current challenges, to name just a few examples, are to improve the spatial and temporal resolution of the quantities of interest (e.g., proteins) and to access quantitative information such as local densities, reaction rates, transport properties, and forces.
In addition, conducting experiments under well controlled conditions, where only one or a few parameters are adjusted at a time, is often difficult owing to the associated technical demands, as well as the inherent complexity of biological systems.
Future progress in this area would greatly enhance our ability to make more detailed comparisons with theory. 
 
Concerning computational approaches, the simulation of multiphysics problems presents a major obstacle.
In particular, the numerical implementation of bulk-boundary coupled reaction-diffusion systems in combination with hydrodynamics and deformable, time-evolving membranes, is an important task for future research.
The primary difficulties here lie in the development of an efficient and stable numerical approach that allows one to solve multiphysics problems in which the numerical domain itself is part of the solution.
In the case of reaction-diffusion dynamics on dynamic membranes without coupling to a bulk volume, this can be addressed by deriving the time-evolution of the surface from the (normal) variation of a free energy functional that describes the mechanical properties of the membrane~\cite{Rangamani.etal2014,Wu.etal2018,Mietke.etal2018,Alimohamadi.Rangamani2018, Mietke.etal2019, Mahapatra.etal2021}.
However, this does not account for dynamics in the bulk, such as intracellular flows and bulk-boundary coupling of protein reactions.
Promising approaches that can cope with these problems in the future are the \emph{level-set} and the \emph{phase-field} methods~\cite{Bray1994, Raetz.Voigt2006}.
These strategies allow one to segregate the computational domain into different regions (e.g. interior and exterior of a cell), where the interface between these regions corresponds to a (smooth) boundary (that could represent, e.g., the cell membrane).
In this way, one can define and solve a coupled set of partial differential equations between different regions, including the interface, and at the same time allow these regions to evolve over time by solving the level-set or phase-field equation.
Most notably, the phase-field method is being used in current research to model cell migration~\cite{Winkler.etal2019}, with applications to reaction-diffusion systems arising only recently~\cite{Marth.Voigt2014, Camley.etal2017, Wang.etal2017, Strychalski.etal2010}.
At the same time, new methods are being developed~\cite{Drawert.etal2016}.
In the long run, it will be a challenge to not only model a deformable domain, but also incorporate the biochemical and mechanical details of membranes in computational approaches.

\section{Summary}
We have presented a summary of the recent progress in understanding the biophysical mechanisms underlying the guidance and control of protein patterns.
In essence, one distinguishes between geometric, biochemical, and mechanical guidance cues.

First, geometric effects can control protein pattern formation, with the cell size affecting the bulk-boundary ratio and the relative penetration depth of cytosolic concentration gradients.
In addition, pattern formation can be limited by finite-size effects.
Geometric effects imposed by the cell shape -- such as the local membrane curvature that controls the distribution of curvature-sensing proteins, and the overall cell shape, which affects the curvature-dependent probability that a protein will encounter the membrane -- can also serve as guiding cues.
Second, we reviewed how protein patterns can guide other protein patterns via biochemical interactions.
Spatial information that is encoded in one protein pattern can be interpreted through protein-protein interactions, thereby transforming the spatial coordinate into a control parameter for downstream protein reactions.
This gives rise to a wide range of different pattern guidance mechanisms, including threshold localization, edge-sensing, and phoretic transport.
Third, mechanical guiding cues, among which flow and stress gradients are of particular relevance, can affect protein pattern formation.
Finally, we outlined open questions and the associated experimental, theoretical, and numerical challenges that need to be faced to improve our understanding of guided pattern formation.

We believe that the mechanisms presented in this review can be applied to a wide range of processes in which spatial information is processed, such as cell migration, cytokinesis, and morphogenesis. 
To advance our understanding of the physical basis and biological relevance of pattern formation, further research on the concepts of pattern guidance will be required, as well as more refined methods to explain experimental observations.
Taken together, this could ultimately contribute to the characterization of general biophysical principles of spatial information processing in living cells.

%
%
%
%

\section*{Supplementary Information}%
\begin{figure*}
    \centering
    \includegraphics[width =0.83\textwidth]{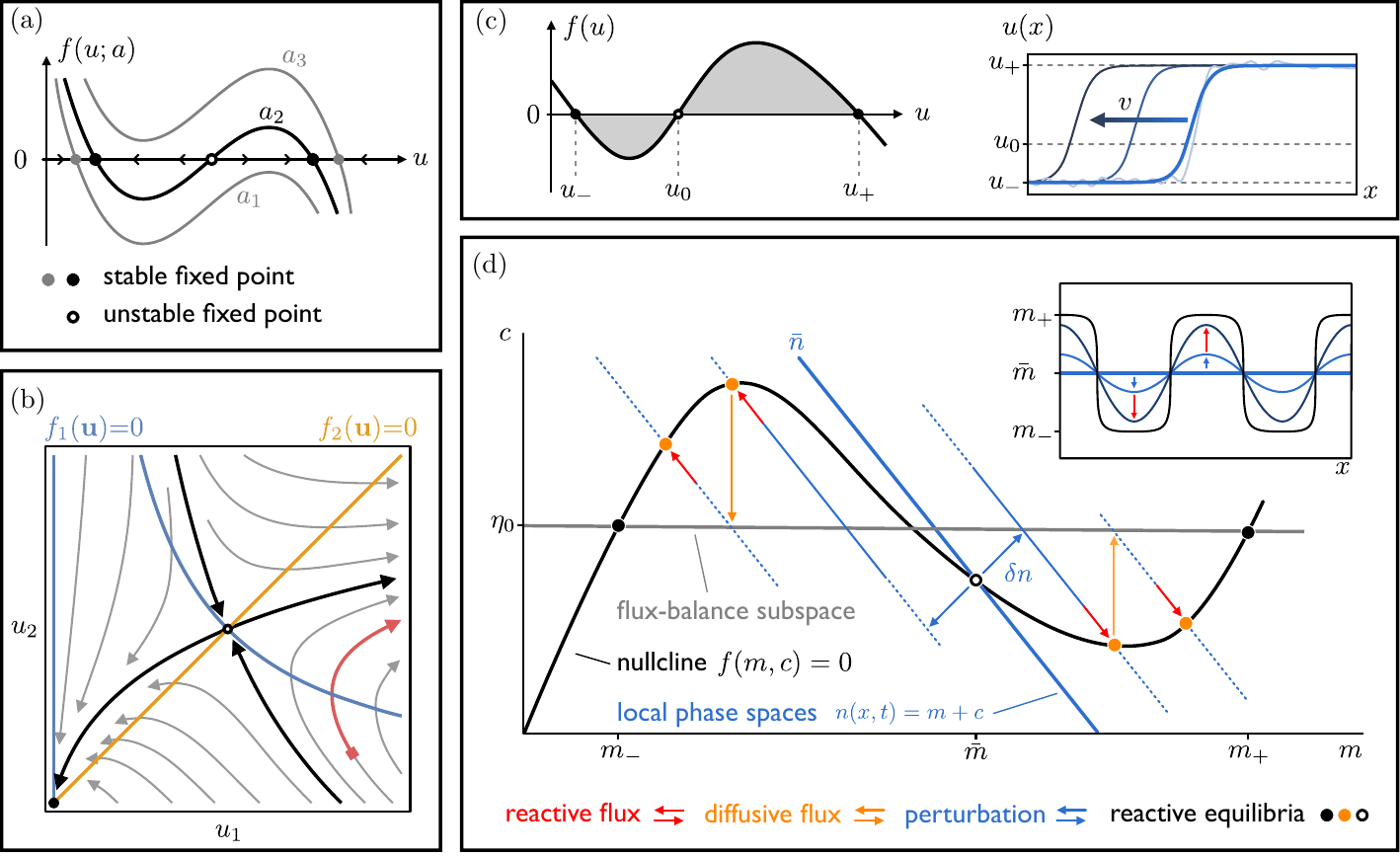}
    \caption{\textbf{Geometric analysis of nonlinear dynamics:}
    (a) The number and stability of reactive equilibria (fixed points) depends, in general, on the reaction kinetics $f(u; a)$, where $a$ is a control parameter.
    (b) Characteristic phase space diagram showing the system's fixed points, which can be derived from the nullclines (blue, orange); the separatrices (black) divide the phase space into qualitatively distinct areas.
    The time evolution of a given initial state (red square) is represented by the flow line associated with this state (red line).
    (c) Front propagation. Left: The reaction kinetics $f(u)$ determine the speed $v$ of the wave. Right: For noisy initial conditions interpolating between the two stable plateaus $u_\pm$, the reaction kinetics first lead to a smoothening of the perturbation and then result in directed front propagation at velocity $v$.
    (d) Illustration of the mass-redistribution instability in phase-space for the biologically relevant limit $D_c \gg D_m$. 
    The homogeneous steady state (black open circle) is determined from the intersection between the local phase space of the total average mass $\bar{n}$ (thick blue line) and the reactive nullcline $f(m,c)=0$ (thick black line).  
    A spatial perturbation $\delta n$ around this homogeneous state causes spatial gradients of the local total density in real space (inset top right). 
    In phase space, the perturbation is represented by local phase spaces (thin blue lines) that contain masses that differ from the homogeneous state, and therefore lead to reactive fluxes (red arrows) towards the reactive equilibria (orange filled circles). 
    This leads to a growing inhomogeneous density distribution in real space, which is further amplified by diffusive fluxes (orange arrows).
    Note that, since cytosolic diffusion is much faster than membrane diffusion $D_c \gg D_m$, diffusive fluxes must point along the vertical direction.
    The steady state density distribution in real space is represented by the flux-balance subspace in phase space (thick gray line)~\cite{Brauns.etal2020}.
    The constant $\eta_0$ determines the vertical position of the flux-balance subspace in phase space and can be interpreted as the (spatial) average cytosolic density.
    }
    \label{fig:box1}
\end{figure*}
\subsection{Methods of analysing pattern formation}
\subsubsection{Reactive equilibrium}
Chemical reactions convert reactants to products and vice versa, thus resulting in fluxes.
An equilibrium state is reached if the sum of all fluxes equals zero, which determines the equilibrium concentrations of constituents.
This equilibrium state is commonly referred to as a \textit{reactive equilibrium}, and is generally distinct from a thermodynamic chemical equilibrium because fluxes can originate from non-equilibrium processes (broken detailed balance)~\cite{Beard.Qian2008}.
One example are NTPase cycles, in which proteins detach from the membrane and must undergo a conformational change before they can re-attach.
The reactive equilibrium in this case is given by a balance between reactive fluxes onto and off the membrane.

Mathematically, the reaction kinetics of a well-mixed system are expressed by ordinary differential equations (ODEs)
\begin{equation}
    \partial_t \mathbf u (t) = \mathbf f ( \mathbf u ) \, ,
    \label{eq:box-ODE}
\end{equation}
where $\mathbf f(\mathbf u)$ contains the (nonlinear) interactions between the components of $\mathbf u$ and therefore corresponds to the sum of individual reactive fluxes. 
Formally, a reactive equilibrium conforms to the steady state solution ${\partial_t \mathbf u=0}$ of Eq.~\eqref{eq:box-ODE} and is termed the \emph{fixed point} of the ODE system, i.e. ${\mathbf f(\mathbf u ^*) = 0}$ for steady state solutions $\mathbf u^*$.
In general, the long-term dynamics are governed by \emph{attractors} of the nonlinear system, whose properties are the subject of the field of dynamical systems theory~\cite{Strogatz1994}.

\goodbreak
\subsubsection{Phase space analysis}
To assess the qualitative dynamics of nonlinear dynamics systems, one must often resort to geometric \emph{phase space} analysis (Fig.~\ref{fig:box1}b).
In phase space, each point corresponds to a specific state of the system, with the \emph{phase space flow} tracing out the time evolution of the system.
Next to the flow lines, fixed points (${\mathbf f(\mathbf u^*) = 0}$) and nullclines (${f_i ( \mathbf u) = 0}$) are characteristic features which reflect the topology of phase space. 
In particular, this representation allows one to identify important features of the system, such as steady states or limit cycles.

As a characteristic example, consider the phase space diagram shown in Fig.~\ref{fig:box1}b, for a two-component system whose dynamics are given by ${\partial_t u_1 = -u_1 + u_1^2 \, u_2}$ and ${\partial_t u_2 = u_1 - u_2}$.
Intersections of the nullclines correspond to fixed points, whose stability can be determined by visualizing the phase space flow.
The system at hand possesses one stable fixed point and one saddle fixed point.
Given a specific initial state, the time evolution of this state can be determined by following the flow line, which provides qualitative information about the system's dynamics.

\goodbreak
\subsubsection{Dispersion relation}
In spatially extended systems, patterns typically form when a (spatially homogeneous) steady state is unstable against random spatial perturbations.
The formal way to probe for instabilities is to perform a linear stability analysis: One first expands spatial perturbations in normal modes and then linearizes the dynamics around a spatially homogeneous steady state $\mathbf u^*$. 
From the linearized system, one can determine the \emph{dispersion relation} $\sigma(q_n)$, which relates the growth rate $\sigma$ of perturbations to their respective mode number $q_n$. 
A typical dispersion relation is shown in Fig.~\ref{fig:shape_size}d.  
Positive values of the growth rate indicate that spatial perturbations are amplified and grow exponentially. 
Since the critical mode $q_c$ with the highest growth rate is expected to dominate near onset, this unstable mode sets the characteristic wavelength of the initial pattern. 
However, in general, the dispersion relation only informs about the characteristic length scale of the pattern in the vicinity of the homogeneous steady state~\cite{Turing1952}; the dominant length scale of the final pattern can be quite different.

\subsection{Nonlinear feedback in protein pattern formation}
\subsubsection{Bistability and propagation of bistable fronts}
Feedback loops are ubiquitous in biological systems and essential for many cellular processes~\cite{Tyson.Novak2010, Ferrell.etal2011, Gelens.etal2014}.
For instance, the calcium waves that follow fertilization of an egg are the result of a positive feedback loop in which cytosolic calcium promotes the flow of additional calcium into the cytoplasm~\cite{Gelens.etal2014}.
In general, feedback loops lead to nonlinear dynamics that exhibit multiple (linearly stable) reactive equilibria~\cite{Cinquin.Demongeot2002}.
A common case is bistability, where the dynamics ${\partial_t u = f(u)}$ has three reactive equilibria, two of which are (linearly) stable ($u_\pm$) and one of which is (linearly) unstable ($u_0$).
Consider a spatially extended bistable system with spatially uniform reaction kinetics $f(u)$, described by the reaction-diffusion equation
\begin{equation}
    \partial_t u(x,t) = D \partial_x^2 u(x,t) + f(u(x,t)) \, .
\end{equation}
In such a system, a front-like profile, where an interface connects two plateaus at the two linearly stable fixed points $u_-$ and $u_+$ (Fig.~\ref{fig:box1}c), will propagate~\cite{Saarloos2003}:
one plateau invades the other with a constant velocity ${v \sim - \int_{u_-}^{u_+} \mathrm d u \: f(u)}$.
These fronts will come to a halt only for a certain choice of parameters, namely when the areas enclosed by $f(u)$ in the intervals $[u_-, u_0]$ and $[u_0, u_+]$ are equal~\cite{Frey.Brauns2020}.

\subsubsection{Mass-redistribution instability}
A general design feature of biochemical networks underlying protein self-assembly is that their dynamics (approximately) preserve the mass of each protein species; i.e., on the time scale of pattern formation, both protein production and protein degradation can be neglected.
Some key features of the patterning dynamics can already be seen with a two-component, mass-conserving system consisting of a cytosolic ($c$) and a membrane ($m$) species in one spatial dimension~\cite{Frey.Brauns2020, Brauns.etal2020}:
\begin{subequations}
\label{eq:box1-2comp}
\begin{align}
\partial_t m(x,t)&=D_m \partial_x^2 m+f(m,c),\label{eq:box1-2comp-a}\\
\partial_t c(x,t)&=D_c \partial_x^2 c-f(m,c)\label{eq:box1-2comp-b}.
\end{align}
\end{subequations}
It is instructive to consider the system's dynamics in $(m,c)$ phase space. 
The \emph{reactive nullcline} (${f(m,c)=0}$) typically shows a N-shape. 
Since the reaction kinetics are mass-conserving, reactive flows tend to remain within the corresponding local phase spaces (${n(x,t)=m+c}$),
and point towards the reactive equilibria determined by the intersection points of these local phase spaces with the reactive nullcline~\cite{Brauns.etal2020}.
Now consider a homogeneous steady state $\bar{n}$ in phase space that intersects the nullcline in a region of negative slope.
Spatial perturbations $\delta n$ around the homogeneous steady state lead to a shift of the local reactive equilibria.
Due to the resulting reactive currents, an upward shift $\delta n$ in total density leads to a decrease in cytosolic density and vice versa (Fig.~\ref{fig:box1}d).
This gives rise to cytosolic concentration gradients, which in turn lead to diffusive fluxes, creating a positive feedback loop.
Eventually, a steady-state pattern is reached when the diffusion currents at the membrane and in the cytosol balance out.
In phase space, the steady state is represented by a \emph{flux-balance subspace} given by ${\widetilde{c}(x)+D_m/D_c\,\widetilde{m}(x)= \eta_0}$, where $\eta_0$ is a constant.
In summary, this pattern formation mechanism involves an intricate coupling between mass-redistribution and local reaction kinetics~\cite{Frey.Brauns2020, Brauns.etal2020}.

\end{document}